\documentclass[acmtog,journal]{acmart}

\usepackage{booktabs} 

\setcitestyle{square}
\usepackage{microtype} 
\usepackage{breakcites}
\usepackage{ellipsis} 
\usepackage[mathletters]{ucs}
\usepackage[utf8x]{inputenc}
\usepackage{mathtools}
\usepackage{amsbsy}
\usepackage{upgreek}
\usepackage{dblfloatfix}

\usepackage[table]{xcolor}

\usepackage{tabularx}
\usepackage{booktabs}

\usepackage{amsmath}

\usepackage{amsfonts}
\usepackage{amssymb}

\definecolor{lightbluishgrey}{rgb}{0.78,0.86,0.93}

\usepackage{layouts}

\usepackage{wrapfig}

\newcommand{\refequ}[1] {Equation~(\ref{equ:#1})}
\newcommand{\refequshort}[1] {Eq.~(\ref{equ:#1})}
\newcommand{\reffig}[1] {Fig.~\ref{fig:#1}}

\makeatletter
\def\reffig{\@ifnextchar[{\@myreffigloc}{\@myreffignoloc}}
\def\@myreffigloc[#1]#2{Fig.~\ref{fig:#2}, \emph{#1}}
\def\@myreffignoloc#1{Fig.~\ref{fig:#1}}
\makeatother

\newcommand{\refsec}[1] {Section~\ref{sec:#1}}
\newcommand{\refapp}[1] {Appendix~\ref{app:#1}}

\let\mat = \mathbf
\usepackage{xfrac}

\usepackage{amsmath}
\usepackage{amssymb}    
\usepackage{cancel}
\usepackage[T1]{fontenc}

\newcommand{\R}{\mathbb{R}}

\newcommand{\vc}[1]{\mathbf{#1}}

\newcommand{\transpose}{{\mathsf T}}

\newcommand{\dn}[1]{∇ #1 ⋅\n}

\newcommand{\gT}[1]{∇ #1^\transpose}
\newcommand{\defon}[2]{\left.{#1}\right|_{#2}}
\newcommand{\n}{\vc{n}}

\renewcommand{\t}{\vc{t}}
\renewcommand{\u}{\vc{u}}
\newcommand{\vv}{\vc{v}}

\newcommand{\x}{\vc{x}}
\newcommand{\y}{\vc{y}}

\newcommand{\A}{\mat{A}}

\newcommand{\D}{\mat{D}}
\newcommand{\E}{\mat{E}}
\newcommand{\F}{\mat{F}}
\renewcommand{\G}{\mat{G}}
\renewcommand{\H}{\mat{H}}

\newcommand{\K}{\mat{K}}

\renewcommand{\L}{\mat{L}}
\newcommand{\M}{\mat{M}}
\newcommand{\MM}{\tilde{\M}}
\newcommand{\N}{\mat{N}}

\newcommand{\V}{\mat{V}}

\newcommand{\X}{\mat{X}}
\newcommand{\Y}{\mat{Y}}

\usepackage{siunitx}
\usepackage{graphicx}

\usepackage{wrapfig}
\newcommand{\wf}[6]{%
  \begin{wrapfigure}[#1]{#2}{#3}%
  \centering%
  \includegraphics[#4]{#5}
  #6%
\end{wrapfigure}}

\makeatletter
\let\save@mathaccent\mathaccent
\newcommand*\if@single[3]{%
  \setbox0\hbox{${\mathaccent"0362{#1}}^H$}%
  \setbox2\hbox{${\mathaccent"0362{\kern0pt#1}}^H$}%
  \ifdim\ht0=\ht2 #3\else #2\fi
  }
\newcommand*\rel@kern[1]{\kern#1\dimexpr\macc@kerna}
\newcommand*\widebar[1]{\@ifnextchar^{{\wide@bar{#1}{0}}}{\wide@bar{#1}{1}}}
\newcommand*\wide@bar[2]{\if@single{#1}{\wide@bar@{#1}{#2}{1}}{\wide@bar@{#1}{#2}{2}}}
\newcommand*\wide@bar@[3]{%
  \begingroup
  \def\mathaccent##1##2{%
    \let\mathaccent\save@mathaccent
    \if#32 \let\macc@nucleus\first@char \fi
    \setbox\z@\hbox{$\macc@style{\macc@nucleus}_{}$}%
    \setbox\tw@\hbox{$\macc@style{\macc@nucleus}{}_{}$}%
    \dimen@\wd\tw@
    \advance\dimen@-\wd\z@
    \divide\dimen@ 3
    \@tempdima\wd\tw@
    \advance\@tempdima-\scriptspace
    \divide\@tempdima 10
    \advance\dimen@-\@tempdima
    \ifdim\dimen@>\z@ \dimen@0pt\fi
    \rel@kern{0.6}\kern-\dimen@
    \if#31
      \overline{\rel@kern{-0.6}\kern\dimen@\macc@nucleus\rel@kern{0.4}\kern\dimen@}%
      \advance\dimen@0.4\dimexpr\macc@kerna
      \let\final@kern#2%
      \ifdim\dimen@<\z@ \let\final@kern1\fi
      \if\final@kern1 \kern-\dimen@\fi
    \else
      \overline{\rel@kern{-0.6}\kern\dimen@#1}%
    \fi
  }%
  \macc@depth\@ne
  \let\math@bgroup\@empty \let\math@egroup\macc@set@skewchar
  \mathsurround\z@ \frozen@everymath{\mathgroup\macc@group\relax}%
  \macc@set@skewchar\relax
  \let\mathaccentV\macc@nested@a
  \if#31
    \macc@nested@a\relax111{#1}%
  \else
    \def\gobble@till@marker##1\endmarker{}%
    \futurelet\first@char\gobble@till@marker#1\endmarker
    \ifcat\noexpand\first@char A\else
      \def\first@char{}%
    \fi
    \macc@nested@a\relax111{\first@char}%
  \fi
  \endgroup
}
\makeatother

\usepackage[ruled,vlined]{algorithm2e}
\usepackage{etoolbox}
\usepackage{array}

\newcommand{\figs}{}
\def\figs/{figs/}

\newcommand{\dx}{\;dx}
\newcommand{\ds}{\;ds}
\newcommand{\dA}{\;dA}
\DeclareMathOperator{\tr}{tr}

\DeclareUnicodeCharacter{8214}{\|}
\DeclareUnicodeCharacter{188}{\tfrac{1}{4}}
\DeclareUnicodeCharacter{189}{\tfrac{1}{2}}
\usepackage[verbose]{newunicodechar}
\DeclareUnicodeCharacter{0923}{\boldsymbol{\Lambda}}

\begin{document}


\title{
Natural~Boundary~Conditions  
for~Smoothing~in~Geometry~Processing}
\acmSubmissionID{0427}

\author{Oded Stein}
\affiliation{Columbia University}

\author{Eitan Grinspun}
\affiliation{Columbia University}

\author{Max Wardetzky}
\affiliation{Universit\"at G\"ottingen}

\author{Alec Jacobson}
\affiliation{ETH Zurich, Columbia University, University of Toronto}


\begin{abstract}
In geometry processing, smoothness energies are commonly used to model
scattered data interpolation, dense data denoising, and regularization during
shape optimization.
The squared Laplacian energy is a popular choice of energy and has a
corresponding standard implementation: squaring the discrete Laplacian matrix.
For compact domains, when values along the boundary are not known in advance,
this construction \emph{bakes in} low-order boundary conditions.
This causes the geometric shape of the boundary to strongly bias the solution.
For many applications, this is undesirable.
Instead, we propose using the squared Frobenious norm of the Hessian as a
smoothness energy.
Unlike the squared Laplacian energy, this energy's \emph{natural boundary
conditions} (those that best minimize the energy) correspond to meaningful
high-order boundary conditions.
These boundary conditions model free boundaries where the shape of the boundary
should not bias the solution locally.
Our analysis begins in the smooth setting and concludes with discretizations
using finite-differences on 2D grids or mixed finite elements for triangle
meshes.
We demonstrate the core behavior of the squared Hessian as a smoothness energy
for various tasks.

\end{abstract}

\begin{teaserfigure}
\includegraphics[width=\textwidth]{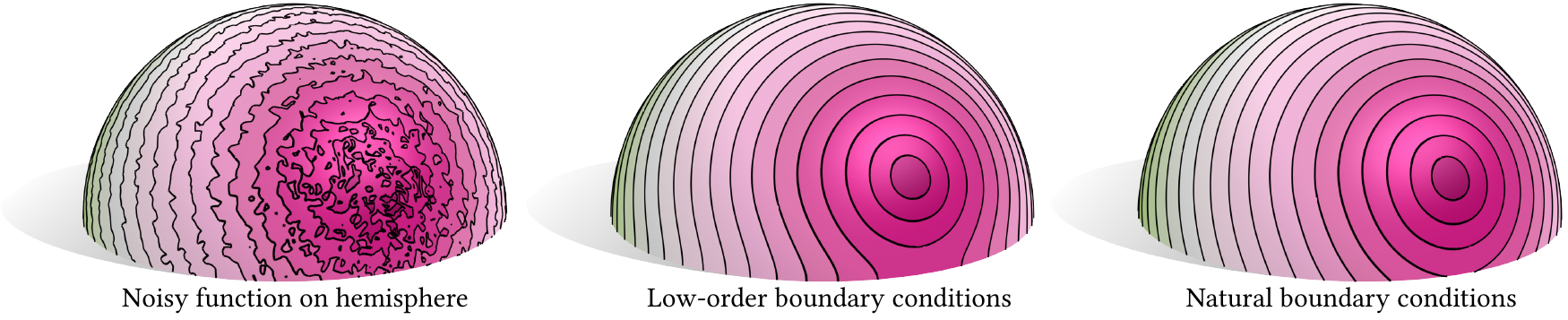}
\caption{Smoothing a noisy function with common low-order boundary conditions
introduces a bias at the boundary: isolines exit perpendicularly. We propose
using a different smoothness energy whose natural boundary conditions avoid this
bias.\label{fig:hemisphere}}
\end{teaserfigure}

\maketitle

%
%
\section{Introduction}

%

%
%

Smoothness energies are commonly used in graphics, geometry processing and
image processing to model deformations, denoise densely sampled
measurements, and interpolate sparse scattered data.
%
Compared to those involving first derivatives, smoothness energies involving
second derivatives are by definition more expressive, but also more challenging
to control.
For example, minimizing the squared-norm of the gradient of an unknown function
$u$ integrated over a bounded domain $Ω⊂\R^2$
\begin{equation}
\label{equ:squared-gradient-energy}
E_{∇²}(u) = ½∫\limits_{Ω} ‖∇u‖² \dA,
\end{equation}
results in a second-order Laplace equation with \emph{only} one set of
prescribable boundary conditions (either fixed values, normal derivatives, or
linear combinations thereof).

Meanwhile, minimizing the \emph{squared Laplacian energy}
\begin{equation}
\label{equ:squared-laplacian-energy}
E_{∆²}(u) = ½∫\limits_{Ω} (∆u)² \dA,
\end{equation}
results in a fourth-order
\emph{bi}-Laplace equation with many different combinations of prescribable boundary
conditions (values and normal derivatives; values and Laplacians; normal
derivatives and Laplacians; etc.). 
This greater expressive power comes with greater responsibility during
modeling.

\begin{figure*}
  \includegraphics[width=\textwidth]{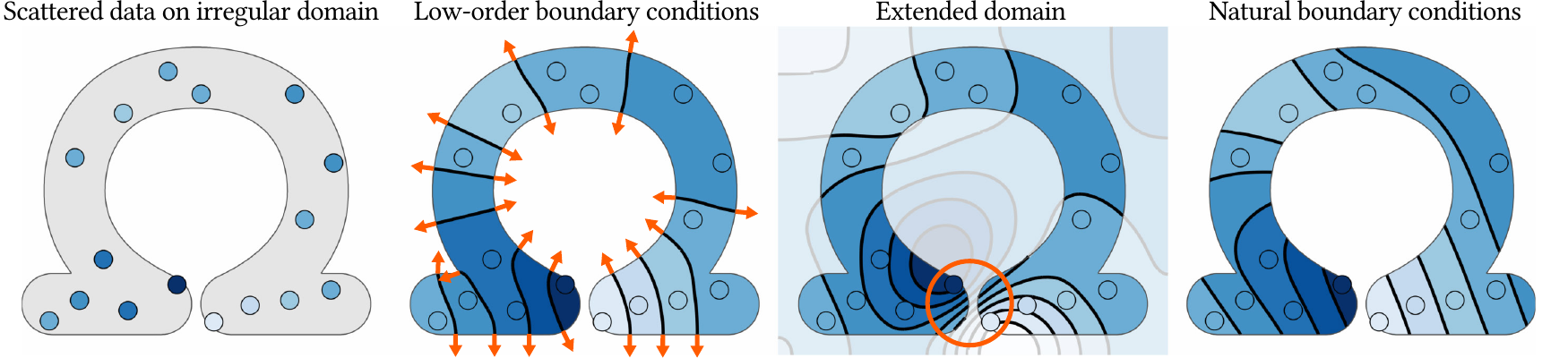}
  \caption{
    When interpolating scattered data over a bounded domain, Neumann conditions
  introduce a noticeable bias at the boundary. Meanwhile, ignoring the boundary
  by interpolating over ambient space allows \emph{bleeding} of data onto
  geodesically distant regions.
  \label{fig:fisherman}}
\end{figure*}

Most previous works enforce low-order boundary conditions. We claim that
these assume a strong \emph{prior} on the unknown function.
For example, consider heat flow over a thermally-conductive plate.
Dirichlet boundary conditions correspond to fixing zeroth-order temperatures
along the plate rim,
whereas Neumann boundary conditions correspond to fixing first-order change in
temperature across the rim into a known medium (e.g., zero change into a
vacuum).

Neumann boundary conditions do not fix values explicitly, so at first glance
these conditions seem appropriate for modeling free boundary problems such as
data smoothing or scattered data interpolation.
However, their effect on the function's first-order behavior near the boundary
leads to noticeable --- often unintentional --- bias (see \reffig{hemisphere}).
Consider the boundary treatment when interpolating scattered data over the 
irregular domain in \reffig{fisherman}.
Standard low-order boundary conditions will cause the interpolation to
\emph{follow} the boundary shape (isolines perpendicular to the boundary,
orange arrows).
These conditions model boundaries that demarcate a salient transition in the
domain (e.g., between a solid object and the vacuum surrounding it).
Sometimes, the boundary simply encloses the region where data is 
defined, without making a strong statement of how space is carved up.
One option would be to \emph{ignore} the given domain and instead solve over
a naive extension into the surrounding space.
As a clear consequence, data will \emph{bleed} across the original domain
boundaries (orange circle).
Natural boundary conditions of a carefully chosen high-order smoothness energy
\emph{model} a different behavior, where boundaries still demarcate the shape
but allow the interpolated data to determine the low-order behavior near the
boundary.
Isolines are not biased toward the boundary and bleeding is avoided.

%

\begin{figure*}[b!]
\includegraphics[width=\linewidth]{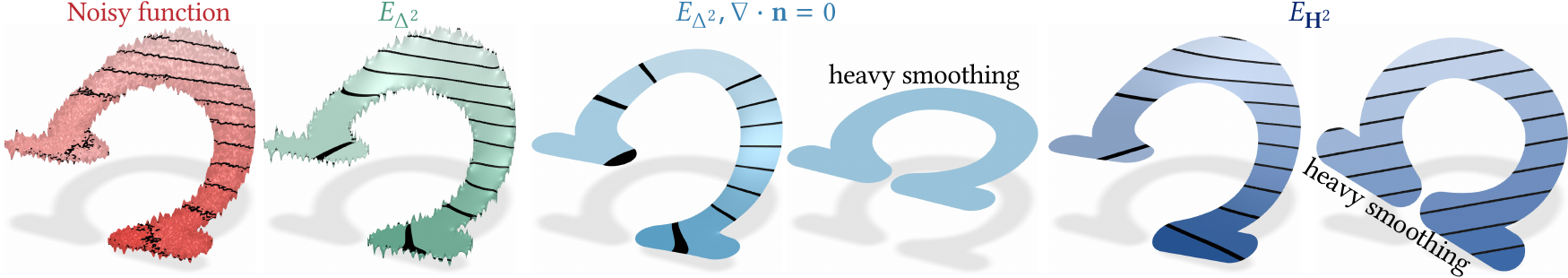}
\caption{
The high-order natural boundary conditions of $E_{∆²}$ maintain noisy
boundaries (green).
Previous works use low-order boundary conditions, but this biases the solution
to \emph{follow} the boundary even if the input data does not (light blue).
Smoothing converges to a constant function.
In contrast, the high-order natural boundary conditions of $E_{\H²}$ produce
smooth boundary values without bias and converges to a \emph{linear} function
(dark blue).
\label{fig:omegasmoothing}}
\end{figure*}

In this paper, we consider a measure of smoothness over irregular domains
subject to high-order boundary conditions.
Intuitively, high-order effects are less noticeable. 
However, high-order boundary conditions will require careful consideration.
Some high-order boundary conditions will result in arbitrarily non-smooth
boundary values (see \reffig{omegasmoothing}).
Rather than hunt for alternative high-order boundary conditions for the popular
squared Laplacian energy, we propose minimizing a different second-order
energy, the squared Frobenious norm of the Hessian or simply \emph{squared
Hessian energy}:
\begin{equation}
\label{equ:squared-hessian-energy}
E_{\H²}(u) = ½∫\limits_Ω ‖\H_u‖²_F \dA,
\end{equation}
where $\H_u ∈ \R^{2 × 2}$ is the symmetric matrix of second-order partial
derivatives.
While this energy appears frequently in the mechanics and image processing
literatures, it has not been applied and analyzed as a smoothness energy for
geometry processing on irregular domains.

%

If both Dirichelet and Neumann boundary conditions are enforced, then
minimizers of $E_{\H²}$ and $E_{∆²}$ will be identical.
%
However, these two energies offer dramatically different behavior for ``free''
or \emph{natural} boundaries where no explicit boundary conditions are
enforced. \emph{Natural boundary conditions} are those that, among all possible
boundary conditions, minimize the given energy.

%

Unlike the natural boundary conditions of the squared Laplacian
energy, those of the squared Hessian energy admit well-behaved solutions
without strong bias near the boundary (see \reffig{omegasmoothing}).
We show the implications of this when modeling smoothness in geometry
processing.
We derive these boundary conditions and analyze their effects in detail.
The advantages of the squared Hessian energy are especially pronounced for
irregular domains (e.g., non-convex domains with arbitrarily shaped,
non-axis-aligned boundaries).

To demonstrate practical results for typical applications of smoothness
energies,
we discretize the squared Hessian energy with natural boundary conditions
following the same steps that have been used to discretize the square Laplacian
energy previously, using finite differences on 2D grids and mixed
finite-elements on triangle meshes.
%

Alongside our core contribution of providing a boundary-insensitive smoothness
energy for geometry processing tasks on irregular domains, we also demonstrate
how our formulation leads to a novel $L_1$-norm minimization for piecewise
planar data reconstruction as well as a smooth understanding of previous
discrete methods for linear subspace design for real-time shape deformation.


\section{Background}
\label{sec:related}

Before discussing related works specifically, we establish concretely what we
mean by \emph{natural boundary conditions}. 

\subsection{Definition: natural boundary conditions}
\label{sec:definition-natural-boundary-conditions}

In the calculus of variations, the term natural boundary conditions refers
specifically to boundary conditions appearing in the Euler-Lagrange equation of
an energy minimization problem that were not imposed beforehand (see
\cite[p.~26]{Gelfand1963} or \cite[p.~34]{Giaquinta96}). This definition is
general in that it applies to any energy or partially constrained energy. Just
like other types of boundary conditions, it is possible to witness natural
boundary conditions on one part of a domain's boundary and explicitly enforce
different boundary conditions on another part. 

Natural boundary conditions achieve the lowest energy among all possible
boundary conditions for a given minimization problem.
Consequently, natural boundary conditions heavily depend on the energy being
minimized. As we will see, the natural boundary conditions of the squared
gradient energy in \refequ{squared-gradient-energy}, the squared Laplacian
energy in \refequ{squared-laplacian-energy}, and the
\refequ{squared-hessian-energy} are all strikingly different.

\subsection{Related Work}

The natural boundary conditions of $E_{∆²}$ and $E_{\H²}$ were considered by
Courant and Hilbert~\shortcite{couranthilbert1924}. Since then,
second-derivative energies have been considered in various fields, including
data, image, and geometry processing. Earlier work also relates to ours in
how the energy, boundary conditions, and domain are chosen, and 
in their application. 



\subsubsection{Geometry Processing}
\label{sec:related-work-geometry-processing}
High-order smoothness energies are a popular and powerful tool.
Discretizations of the squared Laplacian energy have been used for surface
fairing \cite{desbrun99implicitfairing}, smooth surface displacements
\cite{BotschKobbelt04,Sorkine:2004,zhou05lmd,Andrews11}, smooth geodesic distance computation
\cite{Lipman:2010:BD}, data smoothing \cite{Weinkauf2010}, and regularization
during other high-level operations
\cite{huang06subspacemesh,Zhou:2010,cao2014facewarehouse,Cao:2015:RHF,Jones:2016:EPD}.
These and other works \emph{square} the discrete cotangent Laplacian for
triangle meshes \cite{Pinkall:1993:CDM}.
This Laplacian matrix $\L$ is \emph{constructed} assuming zero Neumann boundary
conditions (i.e., for a boundary edge $ij$ with opposite angle $α_{ij}$,
$L_{ij} = \cot α_{ij}$).
Squaring this matrix effectively \emph{bakes in} these conditions. 
Jacobson et al.~\shortcite{Jacobson:MixedFEM:2010} confirm that this
construction agrees with a convergent mixed finite element discretization of
minimizing $E_{∆²}$ \emph{subject to} zero Neumann boundary conditions
($∇u⋅\n=0$).

Squaring $\L$ is an intuitive and easy way to handle unconstrained boundaries
when employing $E_{∆²}$.
However, we show that zero Neumann boundary conditions are \emph{not natural}
for $E_{∆²}$ according to the mathematical definition in
\refsec{definition-natural-boundary-conditions}.
Perhaps surprisingly, the true natural boundary conditions of
$E_{∆²}$ are \emph{not} useful in general (see \reffig{omegasmoothing}).
Meanwhile, the ``not natural'' zero Neumann conditions are clearly quite
useful, as evidenced by the sheer number of geometry processing methods that
impose them and achieve great results for their respective applications
\cite{landreneau2010poisson,Lipman:2010:BD,Weinkauf2010,Kavan:2011:PUF,Jacobson:BBW:2011,finch2011freeform,Rustamov2011multiscale,Jacobson:MONO:2012,weber2012biharmonic,Sykora14}.
%
%
And yet, we will present evidence that the natural boundary conditions of
$E_{\H²}$ can be more useful than either the zero Neumann or natural boundary
conditions of $E_{∆²}$ in contexts where boundaries should minimally influence
the solution.


Natural boundary conditions for energies involving \emph{first} derivatives are
found frequently in geometry processing or related
fields such as computer animation \cite{Batty2010,GNRBfFRS2016}.
When parameterizing surfaces, natural boundary conditions for the
as-conformal-as-possible energy remove the burden of prescribing values for
cut-boundaries
\cite{Desbrun02,Levy:2002:LSC,CohenSteiner02,Mullen:2008:SCP,Springborn:2008:CET}.
The boundary conditions imposed by Desbrun et al.~\shortcite{Desbrun02} have
been applied to \emph{different} energies to which they are no longer natural
per \refsec{definition-natural-boundary-conditions}, though they nonetheless
serve the application \cite{gingold2006dtp}.

Previous works in geometry processing have considered how to discretize the
Hessian of a scalar function defined on a triangulated surface
\cite{Tosun:2007:SOU,de2014discrete}. These methods do not explore 
minimization of $E_{\H²}$ or its natural boundary conditions. 

Wang et al.~\shortcite{Wang2015} square a modified cotangent Laplacian to
design linear subspaces for cartoon animation; they do not explicitly discuss
boundary conditions. We will consider this \emph{discrete} energy and its
relationship to others built using discrete exterior calculus
\cite{fisher2007dtv} and non-conforming edge-based elements \cite{Bergou2006}
in \refsec{lsd}.
We show that --- in contrast to these previous discrete energies --- the null
space of $E_{\H²}$ \emph{contains and only contains} affine functions, the
necessary and sufficient condition for linear subspace design as set forth by
Wang et al.

\begin{figure}
\includegraphics[width=\linewidth]{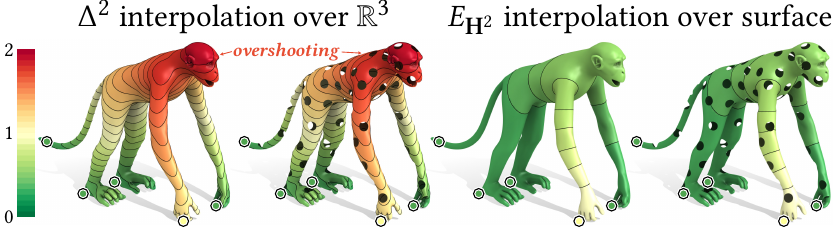}
\caption{Scattered data interpolation using biharmonic radial basis functions
in $\R³$ ignores the embedded surface and therefore also surface boundaries.
In contrast, interpolation by minimizing $E_{\H²}$ on the surface is ``shape
aware'', yet not qualitatively disturbed by open boundaries.
\label{fig:chimp}}
\end{figure}

In the infinite Euclidean domain, biharmonic radial basis functions (a.k.a,
thin-plate splines) minimize $E_{\H²}$ and form a useful basis for various
scattered data interpolation problems \cite{Bookstein:1989:PWT:66131.66134}.
Botsch and Kobbelt~\shortcite{Botsch:2005:RTS} applied these globally supported
functions to shape deformation and modelling by densely sampling selected
surface regions.
Free surface boundaries are not \emph{noticed} any differently from the surface
interior as both are embedded in a deformation of the entire Euclidean space
(see \reffig{chimp}).

\subsubsection{Energies and boundary conditions in other fields}\hfill

Second-derivative smoothness energies such as $E_{∆²}$ and $E_{\H²}$ are found frequently
in denoising, restoration, inpainting, image enhancement, and domain reduction
applications.
%
%
%
Works in image processing that specifically consider
$E_{\H²}$ \cite{terzopoulos1984multilevel,didas2004higher,lysaker2006iterative,roth2009higher,lefkimmiatis2012hessian}
or a broader class of energies including $E_{\H²}$ as a special case
\cite{didas2009,lefkimmiatis2012hessian} impose low-order boundary
conditions, with some notable exceptions: Terzopolous~\shortcite{terzopoulos1984multilevel,terzopoulos1988computation}
was the first to employ the natural boundary conditions of $E_{\H²}$
in the vision and graphics literature, in the context of surface reconstruction from images.
Didas et al.~\shortcite{didas2004higher,didas2009} invoked the natural
conditions for image denoising, whereas Lefkimmiatis et
al.~\shortcite{lefkimmiatis2012hessian} removed boundaries altogether
via periodic or reflexive identification.
When smoothing over the convex image rectangle, free boundaries have
diminished influence on the behavior of the solution.
Applications such as inpainting frequently consider non-convex, irregular image
subregions, but enforce low-order boundary conditions explicitly to ensure
value and derivative continuity \cite{Georgiev04}.

Other works consider related energies, such as the $L_1$-norm of the
Hessian~\cite{lysaker2003noise,steidl2006note,yuan2009total} (a generalization
of \emph{total variation}~\cite{rudin1992nonlinear}), the energy
$E_{∆²}$~\cite{steidl2005relations} or variants thereof~\cite{you2000fourth}.
We show how the $L_1$-norm can analogously apply to smoothing geometric data.

Other authors formulate fourth-order equations directly, not via an energy of
second-derivatives. For instance, Liu et al.~\shortcite{liu2015fourth}
explicitly enforce second- and third-order boundary conditions ($∇u⋅\n = ∇∆u⋅\n
= 0$) on a fourth-order equation similar to the bi-Laplace equation.
As we will see in \refsec{theory}, this set of boundary conditions is not
\emph{natural} to either $E_{∆²}$ or $E_{\H²}$, however they are
frequently imposed on $E_{∆²}$ for geometry processing problems on surfaces
with boundaries.

$E_{∆²}$ and $E_{\H²}$ have been used for the spectral embedding of high
dimensional data~\cite{belkin2004,donoho2003hessian}. A generalization of
$E_{\H²}$ was used by Steinke and Hein~\shortcite{steinke2009non} for
regression, using the energy's natural boundary conditions (therein referred to
as implicit).

\section{One Dimension}
\label{sec:warm-up}
\begin{figure*}
\includegraphics[width=\textwidth]{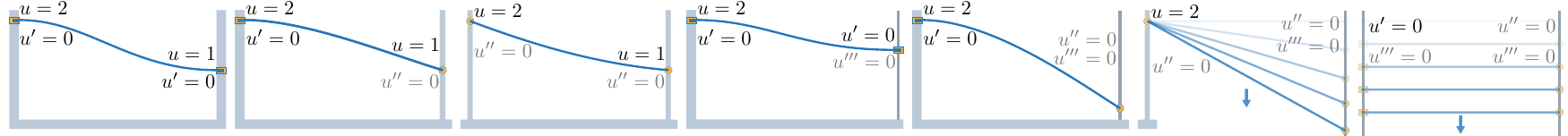}
\caption{A bending bar experiences a uniform load subject to various
  explicit boundary conditions (black, e.g., $u = 2$); in their
  absence, \emph{natural} boundary conditions emerge (gray, e.g.,
  $u'' = 0$). This physical problem has a unique solution when at least two
  low-order boundary conditions are imposed; otherwise, the
  problem becomes underconstrained and the solution tends toward
  infinite values.}
\label{fig:1d-bending}
\end{figure*}

As a didactic exercise, let us derive natural boundary conditions for the
one-dimensional squared second derivative energy integrated over the unit line
segment:
\begin{equation}
\min_u ½∫_0^1 (u'')² \dx,
\end{equation}
where $u' = du/dx$, $u'' = d²u/dx²$ and so on.
Throughout, we appeal to the physical metaphor of a bending bar (see
\reffig{1d-bending}).

In practice, we minimize this energy subject to various boundary conditions or
additional energy terms.
However, applying the calculus of variations to the the energy's raw form will better
illustrate natural boundary conditions.
In the absence of explicit boundary conditions, then $u$ is a minimizer of this
energy if \emph{any} infinitesimal variation $εv$ does not change its energy:
\begin{align}
  0 =  \defon{\frac{d}{d ε}}{ε=0} ½ ∫_0^1 ((u+εv)'')² \dx = ∫_0^1 u''v''\dx  
\end{align}
for all sufficiently smooth functions $v$. In particular, $v$ is
\emph{not} constrained at the boundary.
Integration by parts yields
\begin{align}
  0 = ∫_0^1 u''''v\dx -[u'''v]_0^1+[u''v']_0^1 
\end{align}
for all $v$.
By judiciously testing with $v$ functions, we conclude that $u'''' = 0$ on the
interior interval $(0,1)$ and $u''=u'''=0$ at the boundary $\{0,1\}$.
We began without explicit boundary conditions, so these are the
\emph{natural boundary conditions}.

Natural boundary conditions agree with the energy's null space.
Any linear function $u$ measures zero energy.
Correspondingly, we can interpret the natural boundary conditions as ``the
solution should be linear at the boundary'' ($u'' = 0$). 

We can repeat this process even if we fix \emph{some but not all} boundary
conditions explicitly.
For example, if we fix $u(0)=0$ then the remaining natural boundary conditions
will be $u''(0)=0$  and $u''(1) = u'''(1) = 0$ leaving any linear function
passing through origin as a solution. If we also fix $u'(0)=0$ then the
boundary at $x=0$ is fully constrained and natural boundary conditions only
appear at $x=1$, namely $u''(1)=u'''(1)=0$. In this case, the problem has a
unique albeit trivial solution: $u=0$, the only linear function passing through
the origin with zero slope.

From a mechanics perspective, natural boundary conditions correspond to zero
forces and zero moments at the boundary.
\reffig{1d-bending} shows an example of a uniform load applied to a bending bar
(i.e., $\min_u ∫_0^1 (u'')² + u \dx$) with various explicit boundary conditions
and each problem's resulting set of natural boundary conditions.

%
%

In two dimensions, there will be many different ways to construct a smoothness
energy out of squared second derivatives.
Different energies might even behave the same when boundaries are sufficiently
fixed, but imply \emph{different} natural boundary conditions in the presence
of free boundaries.
Constructing a 2D smoothness energy while only observing behavior on closed
domains or with fully constrained boundaries could lead to surprising behavior
when applied to domains with free boundaries.


\section{Two Dimensions}
\label{sec:theory}

Given a flat 2D domain $Ω ⊂ \R^2$ with a possibly irregular yet smooth
boundary, we consider two intimately related energies that measure 
second-order variations of a function $u: Ω → \R$:
the \emph{squared Laplacian energy} $E_{∆²}(u)$ of
\refequ{squared-laplacian-energy} 
and the \emph{squared Hessian energy} $E_{\H²}(u)$ of
\refequ{squared-hessian-energy}.

\subsection{Green's Identities}
\label{sec:greensidentities}
We will employ various \emph{Green's identities} for functions in \(\mathbb{R}^2\) to move derivatives across
inner products by introducing a boundary term. Starting with the classic
identity:
\begin{equation}
\label{equ:classic-greens}
∫\limits_Ω \left( ∇u ⋅ ∇v  +  u ∆v \right) \dA = ∮\limits_{∂Ω} u\ \dn{v} \ds,
\end{equation}
Replacing $u$ in \refequshort{classic-greens} with $∆u$:
\begin{equation}
\label{equ:greens-with-u-laplace}
∫\limits_Ω \left( ∇∆u ⋅ ∇v  +  ∆u ∆v \right) \dA = ∮\limits_{∂Ω} ∆u\ \dn{v} \ds,
\end{equation}
Replacing $v$ in \refequshort{classic-greens} with $∆v$:
\begin{equation}
\label{equ:greens-with-v-laplace}
∫\limits_Ω \left( ∇u ⋅ ∇∆v  +  u ∆²v \right) \dA = ∮\limits_{∂Ω} u\ \dn{∆v} \ds,
\end{equation}
Replacing $u$ and $v$ in \refequshort{classic-greens} with $\gT{u}$ and $\gT{v}$:
\begin{equation}
\label{equ:greens-with-hessian}
∫\limits_Ω \left( \H_u : \H_v  +  ∇u ⋅ ∇ ∆v \right) \dA = ∮\limits_{∂Ω} \gT{u} \H_v \n \ds,
\end{equation}
Replacing $u$ and $∇v$ in \refequshort{classic-greens} with a vector $\u$
and matrix $\V$:
\begin{equation}
\label{equ:greens-with-tensor}
∫\limits_Ω \left( ∇\u : \V  +  \u⋅(∇⋅\V) \right) \dA = ∮\limits_{∂Ω} \u^\transpose \V \n \ds,
\end{equation}
where $\X : \Y := \tr{(\X^T \Y)}$ computes the Frobenius inner product which
generates the Frobenius norm $\X : \X = ‖\X‖²_F$, and $∇⋅\X$ computes the
matrix divergence of $\X$.

\subsection{Equivalence up to boundary conditions}
\label{sec:equivalence}
We first show that --- regardless of boundary conditions --- minimizers of both
$E_{∆²}$ and $E_{\H²}$ satisfy the biharmonic equation ($∆²u=0$) in the
interior of the domain.
To avoid redundant derivations, we introduce a parameter $α$
(cf.~\cite{couranthilbert1924,terzopoulos1984multilevel}) so that setting $α=1$
gives $E_{\H²}$ and $α=0$ gives $E_{∆²}$:
\begin{align}
\label{equ:parameterized-energy}
½∫\limits_Ω \left((1-α)(∆u)² + α ‖\H_u‖²_F \right) \dA.
\end{align}

As in \refsec{warm-up}, $u$ is a minimizer if adding \emph{any} infinitesimal
variation $εv$ will not change its energy. Thus, 
\begin{align}
\label{equ:variation}
0 = ∫\limits_{Ω} \left( (1-α) ∆ u ∆ v + α \H_u : \H_v \right)\dA
\end{align}
for all $v$. 
Applying Equations (\ref{equ:greens-with-u-laplace}) and
(\ref{equ:greens-with-hessian}) this becomes
\begin{align}
0= &∫\limits_Ω \left(-(1-α)∇∆u ⋅ ∇v  - α ∇v⋅∇∆u\right) \dA\,+ \\
   &∮\limits_{∂Ω} \left((1-α) ∆u\ \dn{v} + α \gT{v} \H_u \n\right)\ds, \notag
\end{align}
for all $v$. Then applying \refequ{greens-with-v-laplace}, we arrive at:
\begin{align}
\label{equ:cov-result}
0= &∫\limits_Ω v∆²u \dA\,+ \\
   &∮\limits_{∂Ω} \left(-v\ \dn{∆u} + (1-α) ∆u\ \dn{v} + α \gT{v} \H_u \n\right)\ds, \notag
\end{align}
for all $v$.
Because this equality must hold for any choice of $v$, 
$u$ must be \emph{biharmonic} in the interior, regardless of 
$α$:
\begin{align}
∆²u = 0 \text{ on $Ω$}.
\end{align}

\begin{figure}
\centering
\includegraphics[width=\linewidth]{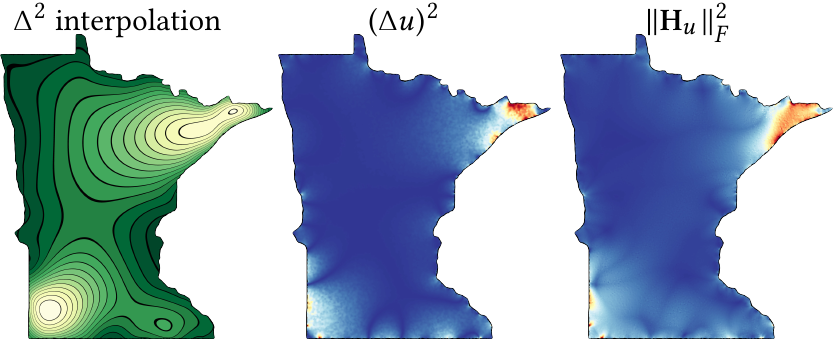}
\caption{
Subject to fixed boundary values and normal derivatives, the same biharmonic
interpolation minimizes \emph{both} $E_{∆²}$ and $E_{\H²}$,
%
despite each energy measuring a different local quantities.
\label{fig:mn}
}
\end{figure}

Biharmonic functions are uniquely determined by boundary values and normal
derivatives \cite[pp.~345]{evans98}, therefore \emph{minimizers} of $E_{∆²}$
and $E_{\H²}$ on \(\mathbb{R}^2\) will be identical when \emph{explicit low-order boundary
conditions} are prescribed, i.e., fixing Dirichlet ($u=f$) as well as Neumann
($\dn{u}=g$) conditions.
Indeed, enforcing such boundary conditions during calculus of variations
requires considering variations such that $v=∇v⋅\n=0$ on the boundary. This
immediately implies that $∇v=0$ on the boundary and correspondingly that the
entire boundary integral in \refequ{cov-result} vanishes.
In conclusion, when modeling smoothness for boundary-value interpolation
problems, these energies are interchangeable despite measuring different local
quantities (see \reffig{mn}).

\begin{figure*}
\includegraphics[width=\linewidth]{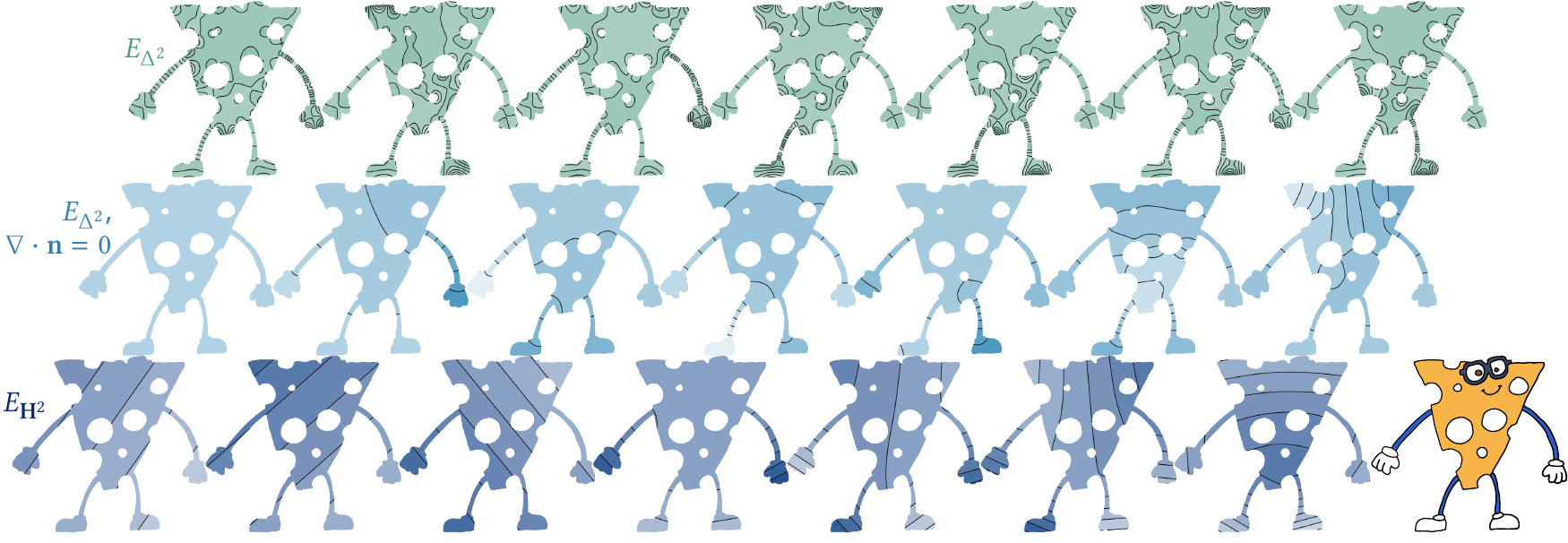}
\caption{
Ten lowest frequency modes of $E_{∆²}$ with natural boundary conditions on a 2D
\emph{Cheeseman} include random harmonic functions with high oscillation near
the boundary (top).
Adding zero Neumann boundary conditions to $E_{∆²}$ causes modes to reproduce
familiar Laplacian eigenfunctions (middle).
In contrast, modes of $E_{\H²}$ with natural boundary conditions include affine
functions and other low frequency functions without constrained normal
derivatives (bottom).
\label{fig:modal-analysis}
}
\end{figure*}

%

\subsection{Natural boundary conditions}
%
%
%
%

In the absence of explicit boundary conditions, the variations $v$ in 
\refequ{cov-result} include all sufficiently smooth functions.
We choose from them specific variations to expose the natural boundary
conditions of the parameterized family of energies in
\refequ{parameterized-energy}.
First, consider all $v$
that vanish on the boundary but whose gradient exists only in the normal
direction ($\defon{v=0 \  \text{and} \  ∇ v = g \n}{∂Ω}$, for some arbitrary
function $g$).
Second-order natural boundary conditions must hold:
\begin{align}
\label{equ:second-order-natural-bcs}
 (1-α)∆u + α\n^\transpose\H_u\n = 0 \ \text{ on $∂Ω$}.
\end{align}
Additionally, considering all $v$ with zero normal derivative without
restricting its value along the boundary (\allowbreak{$\defon{\dn{v} = 0}{∂Ω}$}), we witness
third-order boundary conditions must hold:
\begin{align}
\label{equ:third-order-natural-bcs}
\dn{∆u} + α∇\left(\t^\transpose \H_u \n\right)⋅\t = 0 \text{ on $∂Ω$}.
\end{align}
Both conditions depend on $α$, and therefore the natural boundary conditions
for the squared Laplacian and squared Hessian energies indeed differ (ref.\
\cite{couranthilbert1924}).
We now build an intuition for these conditions for canonical choices of $α$ and
contrast their behavior when modeling smoothness on domains with free
boundaries.

\subsection{Contrasting natural boundary behavior}
%
%
For the squared Laplacian energy $E_{∆²}$ ($α=0$), unconstrained minimizers of
$E_{∆²}$ ($α=0$) will be biharmonic on the interior ($∆²u=0$) and satisfy
natural boundary conditions forcing the solution to be harmonic along the
boundary ($∆u=0$) and continue to be harmonic across the boundary
($\dn{∆u}=0$).
Consistent with our understanding from
\refsec{definition-natural-boundary-conditions}, these boundary conditions
agree with the energy's null space: any harmonic function $u$ measures zero
energy $E_{∆²}(u) = 0$.

\begin{figure} 
\centering
\includegraphics[width=\linewidth]{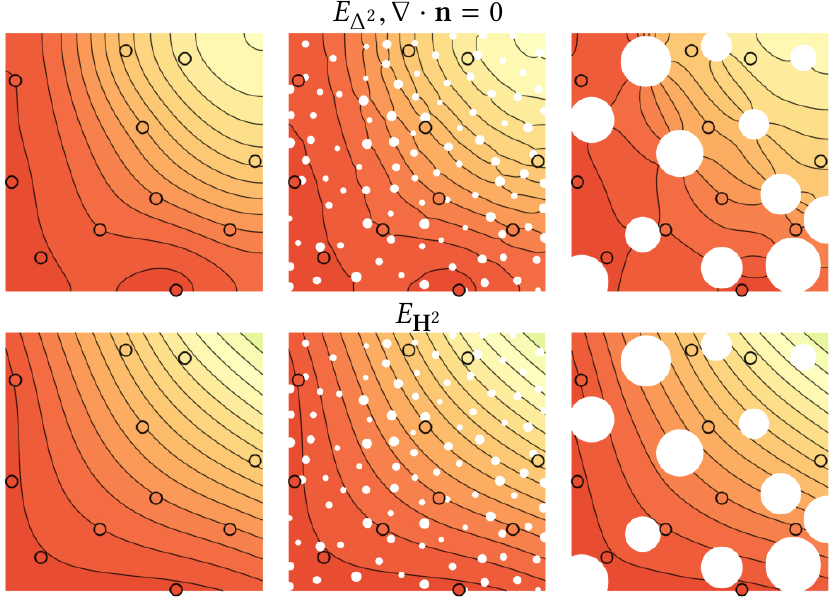}
\caption{Interpolating scattered samples of a quadratic function over a square
domain with (left) and without holes (center and right). When minimizing
$E_{∆²}$ with Neumann boundary conditions, the presence and shape of boundaries
effects the solution. With $E_{\mathbf{\H}²}$, boundaries are less noticeable.
\label{fig:swiss}}
\end{figure}

For the squared Hessian energy $E_{\H²}$ ($α = 1$), unconstrained minimizers
are also biharmonic on the interior ($∆²u=0$), but satisfy \emph{different}
natural boundary conditions forcing the function to be linear in the normal
direction ($\n^\transpose \H_u \n=0$) and after that continue with low variation along the
boundary $\dn{∆u} + ∇\left(\t^\transpose \H_u \n\right)⋅\t = 0$. Again, this
agrees with the null space: all linear functions measure zero energy.

The null space of $E_{∆²}$ is infinite dimensional: for any boundary
values, there exists a harmonic function interpolating them.
This renders the $E_{∆²}$ with natural boundary conditions rather useless as a
smoothness energy.
For data smoothing, noisy data along the boundary will simply remain, since
there exists an interpolating harmonic function that can be added without
affecting the energy (see \reffig[left]{omegasmoothing}). 

In contrast, the linear functions in the null space of $E_{\H²}$ are finite
dimensional and intuitive.
For data smoothing, this null space models ``in the absence of any other
information, fit a linear function'' (see \reffig[right]{omegasmoothing}).

Low-frequency eigenfunctions of smoothness energies  are a standard way to
construct a smooth low-frequency function space in geometry processing.
Typically, previous works use the low-order Dirichlet energy in
\refequ{squared-gradient-energy} with \emph{its} natural boundary conditions
($∇u⋅\n=0$) \cite{zhang07,Hildebrandt:2011:ISM}.
Such modal analysis on the higher-order $E_{∆²}$ with \emph{its} natural
boundary conditions is not practical because
low-frequency modes are polluted with arbitrary harmonic functions (see
\reffig[top]{modal-analysis}).
Explicitly enforcing zero Neumann boundary conditions $∇u⋅\n =0$ on $E_{∆²}$
alleviates this but simply results in the same spectrum as $E_{∇²}$ (up to
squared eigenvalues, \cite{Lipman:2010:BD}).
In contrast, the low frequency modes of $E_{\H²}$ provide a new, smooth, and
well-behaved basis (see \reffig{modal-analysis}).


\subsubsection{Mixing with explicit boundary conditions}
\label{sec:explicit-bcs}
We examine how \emph{explicitly} enforcing certain boundary conditions changes
the remaining natural boundary conditions.
For example, to explicitly enforce values of $u$ along the boundary
(Dirichlet conditions), we must assume that our test function $v$ vanishes
on the boundary ($\defon{v=0}{∂Ω}$ in
Equations~\ref{equ:variation}-\ref{equ:cov-result}. 
\emph{Only}
second-order natural boundary conditions remain. 
For $E_{∆²}$, the additional natural boundary conditions are $∆u=0$. For
$E_{\H²}$, they are $\n^\transpose \H_u \n = 0$.

%

As discussed in \refsec{related}, many previous works in geometry processing
minimize the squared Laplacian energy $E_{∆²}$ subject to zero Neumann
conditions ($\defon{\dn{u}}{∂Ω} = 0$).
We obtain that the \emph{only} additional natural boundary conditions are
third-order: $\dn{∆u} =0$.
%

For data smoothing, zero Neumann boundary conditions encourage functions to be
flat near the boundary.
The null space correspondingly shrinks to contain only constant functions.
%
If the smoothness energy dominates over the data, then a best fit constant
function is found (see \reffig[center]{omegasmoothing}).
%
%
For arbitrary domains, the zero Neumann boundary conditions create a 
strong boundary sensitivity (see \reffig{swiss}).

\subsection{Relationship to previous discrete energies}
\label{sec:lsd}
The null space of the squared Hessian energy $E_{\H²}$ contains \emph{and only
contains} linear functions.
In the context of real-time shape deformation, Wang et al.~\shortcite{Wang2015}
specify that this is precisely the condition needed of a smoothness energy for
linear subspace design.
With this property as a goal, Wang et al.~and other previous works have
designed \emph{discrete} energies by \emph{modifying} the discrete cotangent
Laplace operator, $\L ∈ \R^{n×n}$, for a triangle mesh with $n$ vertices and
$k$ edges.
The unmodified discrete Laplace operator $\L$ can be derived via a
finite-element discretization of the squared gradient energy in
\refequ{squared-gradient-energy}, and the discrete energy preserves the smooth
structure of having only constant functions in its null space. 
In contrast, the following two discrete modifications of $\L$ do not have known
smooth analogs, and as such are more susceptible to errors and more challenging
to analyze \cite{WangError17}.

\begin{figure}
\centering
\includegraphics[width=\linewidth]{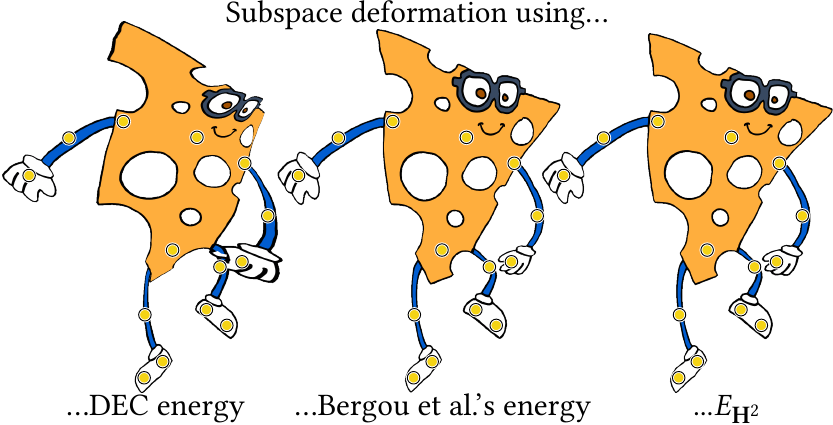}
\caption{The smoothness energy built using discrete exterior calculus in
\refequ{dec-scalar-energy} is not suitable for linear subspace design.
In contrast, deformations using \refequ{bergou-energy} or $E_{\H²}$ are
similarly appropriate.
\label{fig:lsd-dec-vs-ours}
}
\end{figure}

\subsubsection{Discrete exterior calculus}
Fisher et al.~\shortcite{fisher2007dtv} construct a discrete energy for tangent
vector field design using discrete exterior calculus (DEC) on triangle meshes.
Their energy measures the sum of discrete divergences of a given vector field
$\vv$ via:
\begin{equation}
\vv^\transpose \star₁ d₀ \star₀^{-1} \underbrace{d₀^\transpose \star₁}_{\F}\vv,
\end{equation}
where $\star₀$, $\star₁$ and $d₀$ are the discrete Hodge star operators for
zero- and one-forms and the discrete differential matrix for one-forms
respectively, following the adjustments when building $\F$ for free boundaries
as described in \cite{fisher2007dtv}.
If we replace the unknown vector field $\vv$ with the differential of an unknown
scalar field $d₀ \u$, we can build a discrete smoothness energy of a scalar
field:
\begin{equation}
\label{equ:dec-scalar-energy}
\u \ d₀^\transpose \ \F^\transpose \
\underbrace{\star₀^{-1}}_{\M^{-1}} \ 
\underbrace{\F \ d₀}_{\K = \L+\N} \u,
\end{equation}
where the matrix $\M∈\R^{n×n}$ is the (lumped) per-vertex mass matrix, and the
non-symmetric matrix $\K∈\R^{n×n}$ matches the description of the operator in
\cite{Wang2015} constructed by adding the normal derivative operator
$\N∈\R^{n×n}$ to the standard cotangent Laplacian $\L$. This same matrix $\K$
has been previously shown to contain \emph{at least} linear functions in its
right null space \cite{Crane2009}.

\wf{10}{r}{1.01in}{trim=5mm 0mm 0mm 3mm,width=\linewidth}{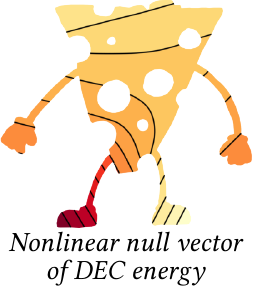}{}
Unfortunately, for some meshes this matrix --- and by extension the 
energy it defines --- contains \emph{other} non-linear functions in its null
space (see inset). This does not appear to be a simpler matter of mesh
resolution or numerics.
From the point of view of linear subspace design (see
\reffig{lsd-dec-vs-ours}), this null space is \emph{too big}.

Following the DEC construction, taking the divergence of the gradient should
correspond to the smooth Laplace operator ($∆ = ∇⋅∇$). If the discrete energy
in \refequ{dec-scalar-energy} were \emph{structure preserving}, then we would
expect it to produce discrete natural boundary conditions for the squared
Laplacian energy $E_{∆²}$ and match its null space of all harmonic functions
(see \reffig{modal-analysis}). However, eigen analysis of this energy for
typical meshes produces a small number ($≥3$) of null modes.
From a structure preservation standpoint, this null space is \emph{too small}.

\subsubsection{Non-conforming finite-elements}
Wang et al.~\shortcite{Wang2015} use an alternative construction to create
their results \cite{WangError17}.
Normal derivatives are added to the edge-based Laplacian $\L_\text{cr} ∈
\R^{k×k}$ resulting from the Crouzeix-Raviart non-conforming finite-element
discretization of the squared gradient energy in
\refequ{squared-gradient-energy}:
\begin{equation}
  \u^\transpose \E^\transpose \K_\text{cr}^\transpose \M_\text{cr}^{-1}
  \underbrace{\K_\text{cr}}_{\L_\text{cr}+\N_\text{cr}} \E \u,
  \label{equ:bergou-energy}
\end{equation}
where $\E ∈ \R^{k×n}$  averages vertex values onto incident edges,
$\M_\text{cr} ∈ \R^{k×k}$ is the Crouzeix-Raviart mass matrix, and
$\N_\text{cr} ∈ \R^{k×k}$ computes normal derivatives at boundary edges.

This energy corresponds to \emph{one of two} discrete energies proposed by Bergou et
al.~\shortcite{Bergou2006} to model plate bending under the assumption of
isometric deformation as the squared Laplacian ($E_{∆²}$) of the displacement
coordinate functions.
Bergou et al.~do not discuss boundary conditions and the discrete boundary
conditions of their two proposed discrete energies \emph{differ} (the other
corresponding to $E_{∆²}$ subject to zero-Neumann boundary conditions).

Perhaps surprisingly, the discrete energy in \refequ{bergou-energy} behaves
very similarly to the squared Hessian energy $E_{\H²}$ despite its motivation
by both Bergou et al.~\shortcite{Bergou2006} and Wang et
al.~\shortcite{Wang2015} as a discrete analog to $E_{∆²}$.
In contrast to the DEC energy in \refequ{dec-scalar-energy}, empirical tests
indicate that the null space of the Bergou et al.~energy contains and only
contains linear functions.
Resulting subspace deformations are visually indistinguishable from those using
$E_{\H²}$ (see \reffig{lsd-dec-vs-ours}).

\begin{figure}
\includegraphics[width=\linewidth]{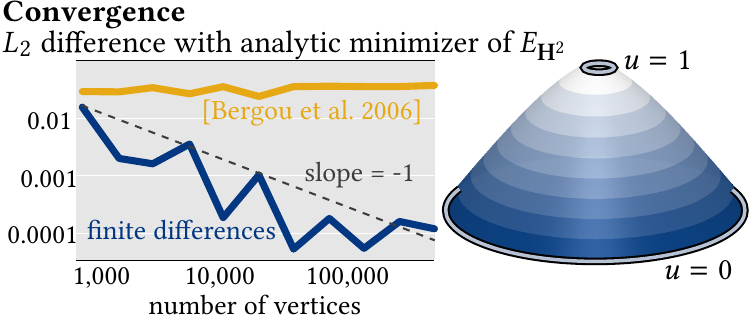}
\caption{On a 2D annulus, fixing inner boundary values to one and outer
boundary values to zero, the analytic minimizer of $E_{\H²}$ is a radially
symmetric function with natural 3rd-order boundary conditions
(see \refsec{explicit-bcs}). While a finite difference discretization of
$E_{\H²}$ converges under refinement, minimizing the discrete energy in
\protect{\cite{Bergou2006}} does not.
\label{fig:annulus-test}}
\end{figure}

Unfortunately, while the discrete energy in \refequ{bergou-energy} clearly does
not behave like a discretization of $E_{∆²}$ with natural boundary
conditions, it also does not converge to smooth minimizers of $E_{\H²}$ with
\emph{its} natural boundary conditions (see \reffig{annulus-test}).
It remains unclear --- and outside our scope --- to show whether this discrete
energy converges to yet some other continuous energy.

\section{Discretization}
\label{sec:discretization}
For our experiments, we discretize the energies $E_{∆²}$ and $E_{\H²}$ using
either the finite difference method on regular 2D grids restricted to a bounded
subregion $Ω ⊂ \R^2$ or the mixed finite element method on irregular triangle
meshes.


\subsection{Finite differences}
For finite differences on a 2D regular grid, we use standard central
differences to compute each entry of the Hessian at all interior nodes
\cite{fornberg1988} (see \refapp{fd}).

Our discretization on grids is in agreement with those used in image processing
on rectangular domains \cite{didas2009}.
As expected, this standard finite difference discretization converges under
refinement (see \reffig{annulus-test}).
No special treatment is required in the case of non-convex, irregular domains
beyond ensuring that all interior nodes have neighboring nodes in all eight
directions.

\subsection{Mixed Finite Elements}
\label{sec:mFEM}
The common discretization of the squared Laplacian energy with zero Neumann
boundary conditions on a mesh with $n$ vertices and $m$ faces is constructed
by \emph{squaring} the cotangent Laplacian $\L ∈\R^{n×n}$:
\begin{equation}
E_{∆²}(u) \defon{\text{subject to } ∇u⋅\n = 0}{∂Ω} ≈ \u^\transpose \L^\transpose \M^{-1} \L \u,
\end{equation}
where $\M ∈\R^{n × n}$ is a (often lumped) \emph{mass} matrix.
Jacobson et al. \shortcite{Jacobson:MixedFEM:2010} demonstrated that this
discrete energy can be derived by applying the mixed finite element method to
the continuous energy minimization problem.

Following their steps, we can similarly discretize the natural boundary
conditions\footnote{Jacobson et al. \shortcite{Jacobson:MixedFEM:2010} mention
``natural'' boundary conditions for $E_{∆²}$ but do not explicitly derive or
discretize them. The free boundaries in their results appear to enforce zero
Neumann boundary conditions.} of the squared Laplacian energy:
\begin{equation}
E_{∆²}(u) ≈ \u^\transpose \L(i,a)^\transpose \M(i,i)^{-1} \L(i,a) \u,
\end{equation}
where $\X(j,k)$ indicates \emph{slicing} rows and columns corresponding to
vertex lists $j$ and $k$ respectively, and specifically $i$ and $a$ are
the lists of interior vertices and all vertices respectively.
It is clear from this construction that \emph{any} discrete harmonic function
($\L(i,a) \u = 0$) will measure zero energy regardless of boundary values. This
is correct from the point of view of structure preservation, and validates the
arbitrary boundary behavior observed when using these natural boundary
conditions for smoothing (see \reffig{omegasmoothing}).

We can follow similar steps to use the mixed finite element method to
discretize the squared Hessian energy. Beginning in the smooth setting, we
introduce an auxiliary matrix-valued variable $\V$ constrained to be equal to
the Hessian of the unknown function on the interior of the domain via a
matrix-valued Lagrange multiplier function $Λ$ which is free on the interior
and clamped to zero on the boundary:
\begin{equation*}
\mathop{\text{saddle}}_{u,\V,Λ} ∫\limits_{Ω} \left(½ \V : \V + Λ:(\H_u-\V)
\right)\dA,
\quad \text{s.t. } \defon{Λ=0}{∂Ω}.
\end{equation*}
Applying the Green's identity from \refequ{greens-with-tensor} to the $Λ:\H_u$
term, we \emph{move} a derivative from $u$ to $Λ$: 
\begin{align}
\mathop{\text{saddle}}_{u,\V,Λ} & ∫\limits_{Ω} \left(½ \V : \V - (∇⋅Λ)⋅∇u - Λ:\V 
\right)\dA
\ + \\
& \cancelto{\defon{Λ=0}{∂Ω}}{∮\limits_{∂Ω}n^\transpose Λ ∇u \ds}. \notag
\end{align}
Applying a functional variation to $\V$, we immediately see that the solution
must be obtained when $\defon{\V=Λ}{Ω}$, so we can substitute $Λ$ out:
%

%
\begin{equation}
\label{equ:mixed-fem-before-discretization}
\mathop{\text{saddle}}_{u,\V} ∫\limits_{Ω} \left(-½ \V : \V - (∇⋅\V)⋅∇u
\right)\dA
\end{equation}
This saddle problem involves only first derivatives. We may use standard
piecewise-linear elements for $u$ and each of the four coordinate functions of
$\V$ (with $\defon{\V=0}{∂Ω}$ because $\defon{Λ=0}{∂Ω}$).
Factoring out the degrees of freedom corresponding to $\V$, we have a
discretization of the squared Hessian energy with natural boundary conditions:
\begin{equation}
\label{equ:mfem-hessian}
E_{\H^2}(u) ≈ \u^\transpose \G^\transpose \A \D \MM^{-1}
\D^\transpose \A \G \u,
\end{equation}
where $\G$,$\A$,$\D$, and $\MM$ are the discrete gradient operator, diagonal
matrix of triangle areas, discrete matrix divergence operator and discrete mass
matrix (see \refapp{mfem}).


\begin{figure}
 \centering
  \includegraphics[width=\linewidth]{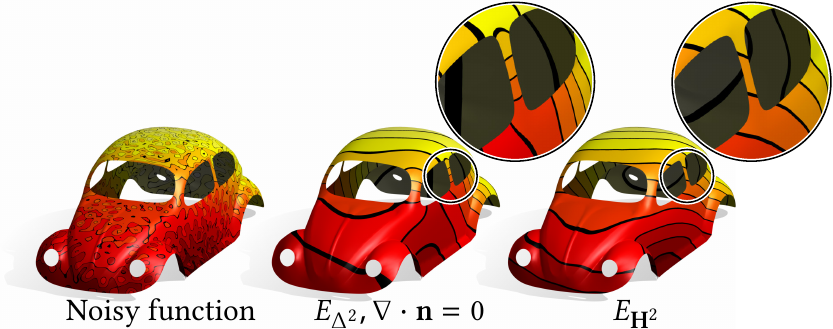}
  \caption{Low-order boundary conditions are the \emph{de facto} standard in
  geometry processing (middle), but they introduce bias at the boundary
  (inset). Switching to the squared Hessian energy alleviates this.\label{fig:beetle}}
\end{figure}

\subsection{Triangle meshes in 3D}
\label{sec:3d}
So far we have only consider flat domains $Ω ⊂ \R²$.
In geometry processing, curved surfaces, and especially those with boundaries,
will benefit analogously from our analysis of smoothness energies (see
\reffig{beetle}). 
We can trivially extend our discrete construction of $E_{\H²}$ to triangle meshes
immersed in $\R²$ by extending the gradient and matrix divergence operators in
\refequ{mfem-hessian} to compute 3D rather than 2D vectors.
This amounts to temporarily treating the Hessian as a $3×3$ matrix.
This extension is inspired by the construction of the discrete Laplacian for
surfaces built by trivially extending the gradient operator to compute
per-face 3D vectors while maintaining the property that: $\L = \G^\transpose \A
\G$.

\subsubsection{Future work: accounting for curvature}\hfill
\label{sec:limitations}
\wf{10}{r}{1.2in}{trim=5mm 0mm 0mm 3mm,width=\linewidth}{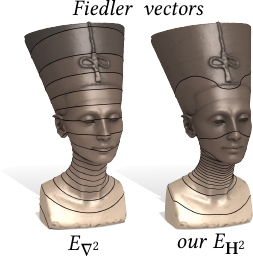}{}
On a closed 
surface, the eigenmodes of the smooth biharmonic equation $∆²u=0$
should match those of the Laplace equation $∆u = 0$ (up to squared
eigenvalues).
For \emph{curved} surfaces, this is not true of our mixed finite element
discretization of $E_{\H²}$ extended to 3D (see inset).
On curved surfaces, the Green's identities from \refsec{greensidentities}
become more complicated, since the gradients are replaced with covariant
derivatives containing curvature terms (i.e., Christoffel symbols).
While building discrete analogs of covariant derivatives is an active topic
(e.g., \cite{Azencot2015,Liu2016}), previous works do not specifically consider
$E_{\H²}$ or --- especially --- its natural boundary conditions.
Regardless, our naive extension to 3D enables our main results concerning free
boundaries to operate on surface meshes.
Accounting for curvature terms on triangle meshes (or possibly some
higher-order representation), with and without boundary, remains as future work.


\section{Experiments \& Applications}
\label{sec:experiments-and-applications}
Smoothness energies are a fundamental ingredient in geometry processing
algorithms. We tour applications using the Hessian to define smoothness.
We solve the resulting sparse linear systems using \textsc{MATLAB} and
quadratic programs using \textsc{Mosek} \cite{andersen17}.
We verified that both our discretizations (see \refsec{discretization})
converge toward the analytic minimizer of $E_{\H²}$ on an annular domain with
radially symmetric fixed value and third-order natural
boundary conditions (see, e.g., \reffig{annulus-test}).
The computational complexity of minimizing $E_{\H²}$ is equivalent to minimizing
$E_{∆²}$: in both cases the number of non-zeros per-row is less than or equal
to the size of the corresponding vertex's two-neighborhood.


The simplest demonstration of these energies is to reconstruct a smooth function
while interpolating values at specific points.
For flat domains with interpolated values and normal derivatives along the
boundary, the reconstructions minimizing $E_{∆²}$ and $E_{\H²}$ will agree (see
\reffig{mn}).
If the boundary is left unconstrained or partially constrained then natural boundary
conditions will appear for each respective energy (see \reffig{fisherman}).
In all examples, differences are best identified by observing how isolines meet
with the boundary: zero Neumann boundary conditions cause the isolines to be
perpendicular to the boundary, while high-order natural boundary conditions
do not force this behavior (see \reffig{swiss}).

%

Linear subspace design for cartoon deformation is a special case of scattered
data interpolation. Instead of interpolating colors or temperatures,
\emph{displacements} are interpolated over a 2D cartoon or surface (ref.,
\cite{skinningcourse:2014}).
Minimizing $E_{\H²}$ with its natural boundary conditions and Kronecker delta
values at specified \emph{point handles}, we 
we can define a
linear basis for smooth deformation interpolation displacements at these
points.
Like previous approaches \cite{Wang2015}, the minimizing functions of $E_{\H²}$
are \emph{linearly precise} and therefore form a coordinate system for $\R²$.
They are therefore a form of \emph{cage-free} generalized barycentric
coordinates (``Hessian coordinates'').
In contrast to previous work, like the coordinates of Wang et al., Hessian coordinates are
defined by a smooth energy where it is easy to show that all and only linear
functions exist in its null space (see \reffig{modal-analysis}).
\reffig{lsd-dec-vs-ours} shows that the deformation behavior is similar that of 
Wang et. al. \shortcite{Wang2015}

\begin{figure}
 \centering
  \includegraphics[width=\linewidth]{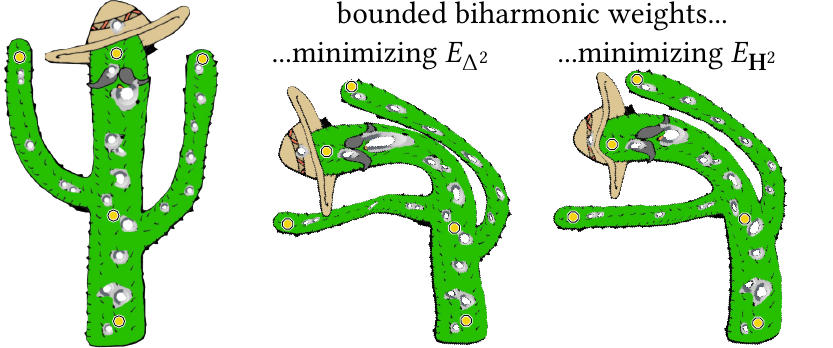}
  \caption{Bullet holes (inner boundaries) cause less distortion when using the
  natural boundary conditions of $E_{\H²}$. \label{fig:cactus}}
\end{figure}

Similarly, automatic methods for computing linear blend skinning weights have
employed smoothness energies \cite{Baran:2007:ARA}.
Classic bounded biharmonic weights minimize $E_{∆²}$ subject to zero Neumann
boundary conditions \cite{Jacobson:BBW:2011} and bound constraints.
Zero Neumann boundary conditions at the outer boundary of a 2D shape give
bounded biharmonic weights their characteristic ``shape-awareness'' property,
but --- as seen in \reffig{cactus} --- the natural boundary conditions of
$E_{\H²}$ are better at \emph{ignoring} the bullet holes shot into the
\emph{Cactus}.

\begin{figure}
 \centering
  \includegraphics[width=\linewidth]{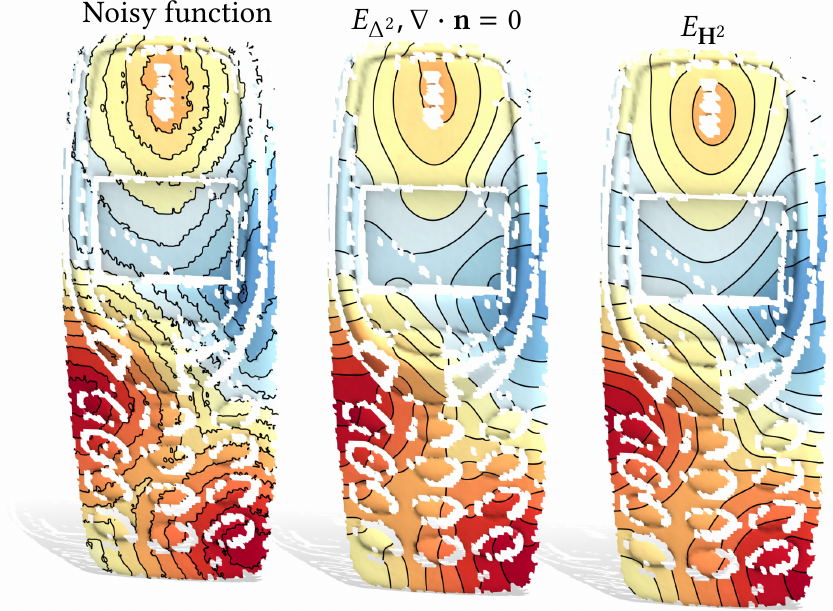}
  \caption{
  Compared to previous methods (left), the natural boundary conditions of
  $E_{\H²}$ (right) better model \emph{ignoring} the non-salient boundaries
  when smoothing data on a partial surface scan.
  \label{fig:phone}}
\end{figure}
\begin{figure}
 \centering
  \includegraphics[width=\linewidth]{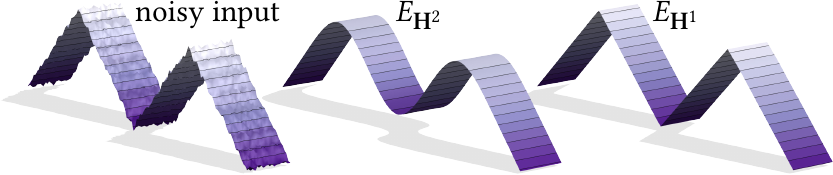}
  \caption{Smoothing using the $L_2$-norm also removes salient creases.
  Minimizing the $L_1$-norm of the Hessian produces a piecewise-planar
  solution. \label{fig:l1vsl2}}
\end{figure}

We now turn our attention to another common use for a smoothness energy: dense
data denoising or fairing.
To do so, we can optimize a function $u$ that minimizes the weighted sum of a
data term ($L_2$-norm of difference between $u$ and a noisy input function $f$)
and the $L_2$ smoothness energy (e.g., $E_{∆²}$ or $E_{\H²}$). 
When using $E_{∆²}$ for data smoothing on domains with free boundaries,
previous methods (e.g., \cite{Weinkauf2010}) enforce low-order boundary
conditions to ensure that noisy boundary values are not simply interpolated
(see \reffig{omegasmoothing}).

The biasing effect of these low-order conditions is apparent: the heavier the
smoothing, the more the solution becomes constant near the boundary
\emph{regardless of the data there}.
In contrast, smoothing with $E_{\H²}$ allows the solution to vary near the
boundary (see \reffig{beetle}).

In \reffig{phone}, we smooth a noisy simulation of heat diffusion on a range
scan of a Nokia cellphone \cite{shrec2009}.
The abundance of free boundaries due to missing data highlights the effect of
zero Neumann boundary conditions compared to the natural boundary conditions of
$E_{\H²}$.

Minimization of an $L_2$-norm --- such as $E_{∆²}$ or $E_{\H²}$ --- prefers to
distribute energy smoothly over the domain.
In contrast, $L_1$ minimization prefers to concentrate high energy at sparse
locations.
At these locations we see the behavior of natural boundary conditions of
the smoothness energy.
Effectively, they become boundaries between low-energy regions.
In \reffig{l1vsl2}, we smooth a toy function (a triangle wave plus noise).
While $L_2$ smoothing $E_{\H²}$ also rounds the peaks, $L_1$ smoothing
$E_{\H¹}$ smooths away the noise but maintains the sharp creases.

\begin{figure}
 \centering
  \includegraphics[width=\linewidth]{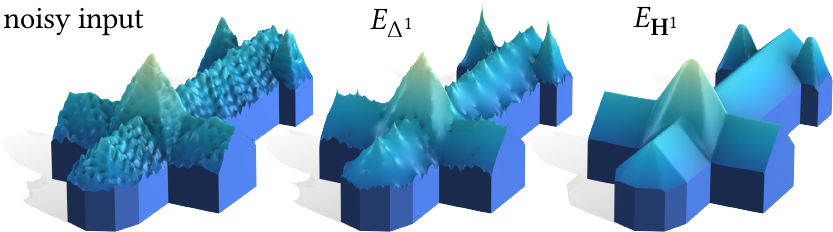}
  \caption{Unlike $L_2$ minimization, the behavior of $E_{∆¹} = ∫ |∆u| \dA$
  and $E_{\H¹} = ∫ |\H_u|_F \dA$ are dramatically different \emph{in the
  interior}.\label{fig:cathedral}}
\end{figure}

In \reffig{cathedral}, we smooth the noisy height data of a cathedral rooftop.
Minimizing the $L_1$-norm of the Laplacian ($E_{∆¹}$)
concentrates energy at isolated points, producing a circus tent appearance.
In contrast, minimizing the $L_1$-norm of the Hessian ($E_{\H¹}$) concentrates
energy along creases, producing a piecewise planar rooftop.

\begin{figure*}
 \centering
  \includegraphics[width=\linewidth]{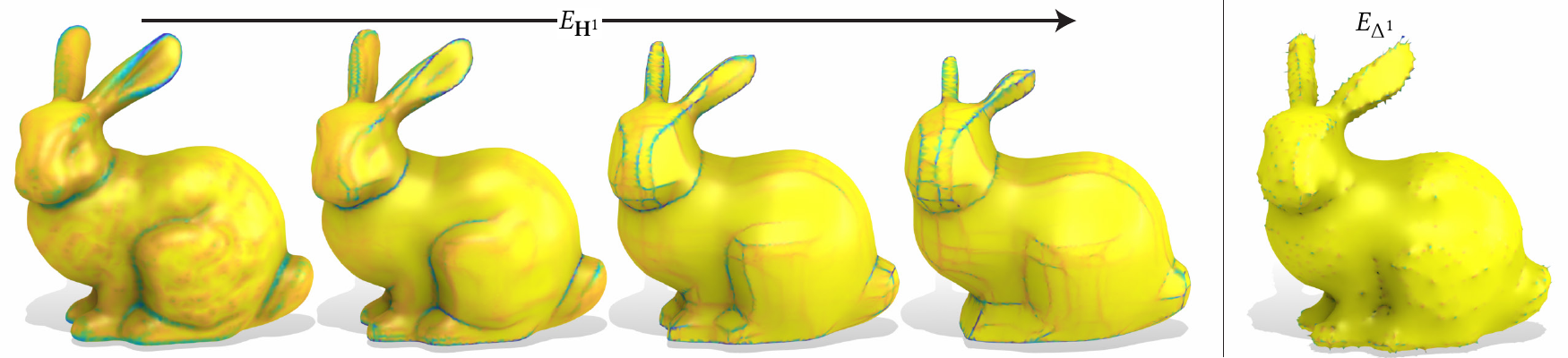}
  \caption{The \emph{Bunny} flows along $L_1$-minimization of the Hessian
  (local energy density in pseudocolor). Energy concentrates at creases and the
  bunny develops smooth, nearly flat regions.
  Minimizing the $L_1$-norm of the Laplacian leads to energy concentration at
  \emph{points} and a prickly appearance.
  \label{fig:l1-bunny}}
\end{figure*}

Under the $L_1$-norm, the difference between minimizers does not rely on the
presence of a boundary.
Indeed, even for closed surfaces with \emph{no} boundary, we see very different
behavior.
In \reffig{l1-bunny}, we treat the surface's geometry as the input data $f$ and
take smoothing steps where the data-versus-smoothness weight acts as an
implicit time-step parameter controlling a geometric flow.
The $E_{\H¹}$ flow of the \emph{Bunny} forms 1D creases bounding smooth,
near-developable (low Gaussian curvature) patches.
This application is inspired by image smoothing with sparse norms
\cite{Xu:2011:ISV} and as such the results are reminiscent of surface contrast
enhancing methods (e.g., \cite{He:2013:MDV}).
In contrast, the $E_{∆¹}$ flow quickly concentrates energy at isolated points
suspended in a near-minimal (low mean curvature) surface.



\section{Conclusion \& future work}
Energies built with the Hessian rather than the Laplacian unlock high-order
boundary conditions that are especially useful when boundaries are to be
conceptually ignored during problem modeling.

In future work, we would like to investigate a discretization of $E_{\H²}$ for
3D triangle meshes that accounts for the covariant derivative (see
\refsec{limitations}).
For flat domains, it would be interesting to explore boundary-only
discretizations (cf.~\cite{Chen:2015:BDH}) or subspace deformation for solid
objects.
Initial derivations suggest that point constraints may not lead to smooth
minimizers of second-order smoothness energies for solids.
One possibility, however daunting, may be to consider the squared Frobenius
norm of the three-tensor of third derivatives, an energy with only quadratic
functions in its null space.

We hope that our work sheds new light on familiar problems and provides
insights into the power of the natural boundary conditions of the squared
Hessian energy for geometry processing problems.
Many applications that are currently using the squared Laplacian energy with
zero Neumann boundary conditions can potentially profit by trying the squared
Hessian energy with natural boundary conditions alongside it.


\bibliographystyle{ACM-Reference-Format}
\bibliography{references}
%

\appendix

\section{Finite Differences}
\label{app:fd}

In order to discretize $E_{\H²}$ over a domain $Ω$ embedded in a grid with $n$
total nodes and $\hat{n}$ interior nodes, we define a sparse matrix $\H =
\left[\H_{xx}^\transpose\ \H_{yy}^\transpose\
\sqrt{2}\H_{xy}^\transpose\right]^\transpose \in \R^{3\hat{n} × n}$ so that
each row of $\H_{xx}$ approximates the second derivative of the unknown
function at the corresponding interior grid node: $\H^i_{xx} \u \approx
∂²u(\x_i)/∂x²$ and analogously for rows in $\H_{yy}$ and $\H_{xy}$.

We use standard, second-order stencils for each term, i.e.,:
\begin{align}
\H_{xx}^i\u                  &= \frac{u_{i  ,j-1} -2 u_{i  ,j}+u_{i  ,j+1}}{h²}\\
\H_{yy}^i\u                  &= \frac{u_{i-1,j  } -2 u_{i  ,j}+u_{i+1,j  }}{h²} \notag\\
\H_{xy}^i\u = \H_{yx}^i\u    &= \frac{u_{i-1,j-1} +u_{i-1,j+1}-u_{i+1,j-1}+u_{i+1,j+1}}{4h²}. \notag
\end{align}

\section{Mixed Finite Elements}
\label{app:mfem}
Solving the piecewise-linear discretization of
\refequ{mixed-fem-before-discretization} by differentiating with respect to all
degrees of freedom, we have a system of linear equations in matrix form:
%
%
\begin{equation}
\left(\begin{array}{cc}
\MM 
  & \G^\transpose 
  \A
  \D \\
\D^\transpose 
  \A
  \G & 0
\end{array}
\right)
\left(\begin{array}{cc}
\V_{xx} \\
\V_{xy} \\
\V_{yx} \\
\V_{yy} \\
\u      
\end{array}
\right)
= 0
\end{equation}
where $\MM ∈ \R^{4|i|×4|i|}$ repeats the mass matrix $\M(i,i)$ along the
diagonal, $\A∈\R^{2m×2m}$ is a diagonal matrix containing triangle areas, and
$\D ∈ \R^{2m×4|i|}$ computes the matrix divergence defined by:
\begin{equation}
\D = 
\left(\begin{array}{cccc}
\G(x,i) & \G(y,i) & 0       & 0 \\
      0 &       0 & \G(x,i) & \G(y,i)
\end{array}\right),
\end{equation}
where $\G ∈ \R^{2m × n}$ is the usual gradient operator and $\G(x,i)$ selects
the rows and columns corresponding to the $x$-components and interior vertices
respectively.
Finally, if we use lumped mass matrices we can efficiently condense this system
and use it to define a discretization of our original energy:
\begin{equation}
E_{\H^2}(u) ≈ \u^\transpose \G^\transpose \A \D \MM^{-1}
\D^\transpose \A \G \u.
\end{equation}

\section{$L_1$ minimization}
\label{app:qp}

In the smooth setting, the \(L^1\) Frobenious norm of the Hessian
is:
\begin{equation}\label{eq:l1smoothlaplacian}
∫\limits_Ω \left| \H_u \right|_F \dA
\end{equation}

This can be minimized by introducing an auxiliary matrix-valued variable equal
to the element-wise absolute value of the Hessian $\Y = |\H_u|$ and solving the
constrained optimization problem:
\begin{align}\label{eq:discl1}
\min_{u,\Y}        &\quad ∫\limits_Ω \mathbf{1}^\transpose \Y \mathbf{1} \dA \\
\text{subject to } &\quad \H_u \leq \Y,\ \H_u \geq -\Y,\text{ and } \Y ≥ 0,
\end{align}
where $\mathbf{1}$ is a vector of ones (with appropriate length).

Using our discrete matrices, this becomes a linear program
\begin{align}\
 \min_{\u,\y}       &\quad  \mathbf{1}^T \MM \y \\
 \text{subject to } &\quad  \H  \u \geq -\y \\
                    &\quad  \H  \u \leq \y \\
                    &\quad  \y \geq 0,
\end{align}
where $\y ∈ \R^{4\hat{n}}$ is a vectorization of per-vertex $2×2$ matrices on a
mesh with $\hat{n}$ interior vertices and $\H = \D^\transpose \A \G$.

When combined with other quadratic energies such as a data-term, this
transforms into a quadratic program. We solve such problems with \textsc{Mosek}
\cite{andersen17}.

\end{document}